\documentclass{SolarPhysics}    
\usepackage[optionalrh]{spr-sola-addons}
\usepackage{epsfig}
\usepackage{url}

\newcommand{\gapprox}{\lower.4ex\hbox{$\;\buildrel >\over{\scriptstyle\sim}\;$}}
\newcommand{\lapprox}{\lower.4ex\hbox{$\;\buildrel <\over{\scriptstyle\sim}\;$}}

\def\etal{{\it et al.}~}
\def\eg{{\it e.g.}~}
\def\ie{{\it i.e.}~}

\begin{document}
\begin{article}
\begin{opening}
\title{ The State of Self-Organized Criticality of the Sun 
	During the Last Three Solar Cycles. II. Theoretical Model }

\author{Markus J. Aschwanden}
\runningauthor{M. Aschwanden}
\runningtitle{Self-Organized Criticality}

\institute{Solar and Astrophysics Laboratory,
	Lockheed Martin Advanced Technology Center, 
        Dept. ADBS, Bldg.252, 3251 Hanover St., Palo Alto, CA 94304, USA; 
        (e-mail: \url{aschwanden@lmsal.com})}

\date{Received 1 July 2010; Accepted ...}

\begin{abstract}
The observed powerlaw distributions of solar flare parameters can be 
interpreted in terms of a nonlinear dissipative system in the state of 
self-organized criticality (SOC). We present a universal analytical model 
of a SOC process that is governed by three conditions:
(i) a multiplicative or exponential growth phase, (ii) a randomly
interrupted termination of the growth phase, and (iii) a linear decay phase. 
This basic concept approximately reproduces the observed frequency
distributions. We generalize it to a randomized exponential-growth model,
which includes also a (log-normal) distribution of threshold energies before 
the instability starts, as well as randomized decay times,
which can reproduce both the observed occurrence frequency distributions
and the scatter of correlated parametyers more realistically. With this 
analytical model we can efficiently perform Monte-Carlo simulations of 
frequency distributions and parameter correlations of SOC processes, which are
simpler and faster than the iterative simulations of cellular automaton
models. Solar cycle modulations of the powerlaw slopes of flare frequency
distributions can be used to diagnose the thresholds and growth rates 
of magnetic instabilities responsible for solar flares.
\end{abstract}


\end{opening}

\section{       Introduction 	}

In this theoretical study we model the occurrence frequency distributions 
and associated correlations of parameters observed in solar flare hard 
X-rays (Paper I; Aschwanden 2010a) 
using the concept of self-organized criticality (for a 
recent textbook, see Aschwanden, 2010b). 
The concept of self-organized criticality (SOC) has been pioneered
by Bak, Tang, and Wiesenfeld (1987), which they briefly summarize
in their abstract: ``We show that certain extended
dissipative dynamical systems naturally evolve into a critical state,
with no characteristic time or length scales. The temporal ``fingerprint''
of the self-organized critical state is the presence of 1/f noise; its
spatial signature is the emergence of scale-invariant (fractal) structure.''
A consequence of the scale-invariant structure is the manifestation of
powerlaw-like distributions of spatial, temporal, and energy parameters,
which became the observational hallmark of SOC processes.

An interpretation of the omnipresent powerlaw distributions of solar-flare
peak fluxes or energies in terms of SOC models was first proposed by
Lu and Hamilton (1991), which they summarize as follows: 
``The solar coronal magnetic field is proposed to be in a
self-organized critical state, thus explaining the observed power-law
dependence of solar-flare-occurrence rate on flare size which extends
over more than five orders of magnitude in peak flux. The physical picture
that arises is that solar flares are avalanches of many small reconnection
events, analogous to avalanches of sand in the models published by Bak and
colleagues in 1987 and 1988. Flares of all sizes are manifestations of the
same physical processes, where the size of a given flare is determined by
the number of elementary reconnection events. The relation between
small-scale processes and the statistics of global-flare properties which
follows from the self-organized magnetic-field configuration provides a way
to learn about the physics of the unobservable small-scale reconnection
processes. A simple lattice-reconnection model is presented which is
consistent with the observed flare statistics.''
The lattice-based computer simulations of next-neighbor interactions is
known as {\sl cellular automaton model} and represents a powerful numerical
model that can reproduce many observed distributions of SOC processes
(\eg, see review by Charbonneau \etal2001). 

An alternative approach to cellular automaton models of SOC processes 
are analytical models, which also can reproduce the observed statistical 
distributions and scaling laws between observed parameters, which
we pursue here. 

\section{Analytical SOC Model}

A simple analytical model of SOC processes with universal applicability
can be constructed from the concept of a nonlinear process that: 
(i) has a multiplicative or exponential growth phase, (ii) is randomly
interrupted, and (iii) has a linear decay phase. The basic evolution of 
such a dissipative nonlinear process with avalanche-like characteristics
is depicted in Figure 1. 

The exponential-growth model with conditions
(i) and (ii), which constrains a powerlaw distribution of event sizes,
goes back to Willis and Yule (1922), who applied it to geographical 
distributions of plants and animals. According to Simkin and
Roychowdhury (2006), Yule's model was re-invented by Fermi (1949),
who applied it to the origin of cosmic rays, and by Huberman and Adamic
(1999), who applied it to the growth dynamics of the World-Wide Web.
In the astrophysical context, it was applied to cosmic ray 
and solar flare transients by Rosner and Vaiana (1978). However,
the waiting time between subsequent events was interpreted as
energy storage time of the exponential energy build-up phase in
the model of Rosner and Vaiana (1978), which was not confirmed
observationally (Lu, 1995; Crosby, 1996; Wheatland, 2000; 
Georgoulis \etal2001). The lack of a correlation between waiting times and 
flare size is not surprising in the concept of SOC models. The original 
SOC model of Bak \etal(1987) assumes that avalanches occur randomly
in time and space without any correlation, and thus a waiting time
interval between two subsequent avalanches refers to two different and
independent locations (except for sympathetic flares), and thus bears
no information on the amount of energy that is released in each
spatially separated avalanche. In an alternative model, the exponential 
rise time $t_S$ is assumed to be related directly to the energy 
release process of an instability during a flare (Aschwanden \etal 1998),
rather than to the much longer energy storage time assumed in Rosner
and Vaiana (1978). 

For the decay process of the instability, a linear decay is found 
approximately to reproduce the observed scaling laws between flare
peak counts and flare durations (Aschwanden 2010b), which constitutes
assumption (iii) of our analytical model. The physical implication of
a linear decay is a constant decay rate after the saturation of the
instability, independent of the size of the event. This linear decay
assumption (iii) has been added to our model mostly on empirical grounds.
Tests with cellular automaton models could corroborate its reality.

\begin{figure}
\centerline{\includegraphics[width=0.70\textwidth]{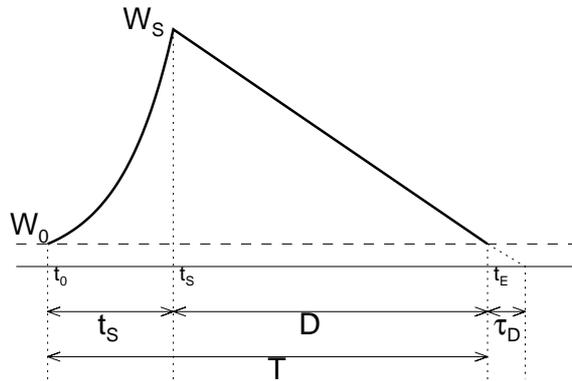}}
\caption{Schematic of the temporal evolution of an avalanche event, consisting
of (i) a rise time ($t_S$) with exponential growth of the energy
release $W(t)$ from a threshold level $W_0$ to the saturation level $W_S$,
and (ii) a decay time ($D$) with a constant decay rate
$\eta = dW/dt = W_0/\tau_D$.}
\end{figure}

\medskip
Let us now formulate this analytical SOC model in mathematical terms.
A full derivation of the exponential-growth model is given in Section
3.1 of Aschwanden (2010b). We define the temporal evolution of 
the energy release rate $W(t)$ of a nonlinear process that starts at a 
threshold energy of $W_0$ with an exponential growth time $\tau_G$. 
The process grows exponentially until it saturates at time 
$t=t_S$ with a saturation energy $W_S$, 
$$
        W_S = W(t=t_S) = W_0 \ \exp{\left({t_S \over \tau_G} \right)} \ .
        \eqno(1)
$$
We define a peak energy release rate $P$ that represents the maximum
energy release rate $W_S$, after subtraction of the threshold energy $W_0$,
$$
	P = W_S - W_0 = W_0 \left[ \exp{ \left( {t_S \over \tau_G} \right)} 
		- 1 \right] \ .
	\eqno(2)
$$
For the saturation times $t_S$,
which we also call ``rise times'', we assume a random probability distribution, 
approximated by an exponential function $N(t_S)$ with e-folding time constant 
$\tau_S$, 
$$
        N(t_S) d t_S = {N_0 \over \tau_S}
	\exp{\left(-{t_S \over \tau_S}\right)} d t_S \ .
        \eqno(3)
$$
This probability distribution is normalized to the total number of events,
$N_0$. The resulting frequency distribution of peak energies, $N(P)$,
is then,
$$
	N(P) dP = 
	N[t_S(P)] \left| {d t_S \over dP} \right| dP =
	{N_0 (\alpha_P - 1) \over W_0} 
	\left({P \over W_0} + 1 \right)^{-\alpha_P} \ dP \ ,
	\eqno(4)
$$
which is an exact powerlaw distribution for large peak energies
($P \gg W_0$) with a powerlaw slope $\alpha_P$ of
$$
	\alpha_P = \left( 1 + {\tau_G \over \tau_S} \right) \ .
	\eqno(5)
$$
The powerlaw slope thus depends on the ratio of the 
e-folding saturation time $\tau_S$ to the
exponential growth time constant $\tau_G$, 
which is essentially the average number of growth times. 

Once an instability has released a maximum amount $W_S$ of energy,
say when an avalanche reaches its largest velocity on a sandpile,
the energy release gradually slows down until the avalanche comes 
to rest. For sake of simplicity we assume a constant energy decay 
rate $\eta$ after the peak of the energy release, lasting for 
a time interval $D$ until it drops to the threshold level $W_0$,
$$
        \eta    = {W_S - W_0 \over D} = {W_0 \over \tau_D}\ ,
        \eqno(6)
$$
which produces a linear decay of the released energy,
$$
        W(t) = W_0 + (W_S - W_0)
        \left( 1 - {(t - t_S) \over D} \right) \quad
        t_S < t < t_E \ ,
        \eqno(7)
$$
where $t_E$ is the end time of the process at $t_E = t_S + D$.
The time interval $D$ of the energy decay thus depends on the peak 
energy release rate $P$. The time interval $T$ of the total duration of the
avalanche process is the sum of the exponential rise phase $t_S$ 
and the linear decay phase $D$ as illustrated in Figure 1, 
$$
	T = t_S + D = \tau_G \ln{\left({P \over W_0} + 1 \right)}
	             + \tau_D \left({P \over W_0}\right) \ .
	\eqno(8)
$$
We see that this relationship predicts an approximate proportionality of
$T \propto P$ for large avalanches, since the second term, which is
linear in $P$, becomes far greater than the first term with a 
logarithmic dependence ($\propto \ln{P}$). 
The resulting distribution of flare durations, $N(T)$, is then
approximately (neglecting the rise time),
$$
	N(T) dT = N[\tau(T)] \left| {d\tau \over dT} \right| dT 
	\approx {N_0 (\alpha_P - 1) \over \tau_D}
	\left( {T \over \tau_D} + 1 \right)^{-\alpha_T} \ ,
	\eqno(9)
$$
which is a powerlaw function for large durations $T$ with the same slope
$\alpha_T=\alpha_P$ as the peak energy rate $P$.

The total released energy $E$ is the time integral of the
energy release rate $W(t)$ during the event duration $T$. Neglecting the
rise time $t_S$ and subtracting the threshold level $W_0$ before the 
avalanche, we obtain
$$
	E = \int_0^{T} \left[ W(t)-W_0 \right] \ dt 
	\approx \int_{t_S}^{t_S+D} \left[ W(t)-W_0 \right] \ dt \ 
	= {1\over 2} P D      \ .
        \eqno(10)
$$
yielding a frequency distribution $N(E)$ of,  
$$
	N(E) dE = N[P(E)] \left| {dP \over dE} \right| dE 
	\approx {N_0 (\alpha_P - 1) \over 2 E_0}  
	\left[ \sqrt{ {E \over E_0} } + 1 \right]^{-\alpha_P}
	\left[ {E \over E_0} \right]^{-1/2} \ .
	\eqno(11)
$$
The resulting frequency distribution $N(E)$ of energies is close to a
powerlaw distribution and converges to the slope $\alpha_E=(\alpha_P+1)/2$
for large energies, 
$$
	N(E) dE \approx
	{N_0 (\alpha_P - 1) \over 2 E_0}   
	\left( {E \over E_0} \right)^{-(\alpha_P+1)/2} \ .
	\eqno(12)
$$
We show the frequency distributions of the total energy $E$, peak energy $P$,
rise time $t_S$, and total duration $T$ in Figure 2, for  
three different ratios of the growth rate to the average saturation time 
$\tau_S$, \ie, $\tau_G/\tau_S$=0.5, 1, and 2. We see that this model can 
accomodate a range of powerlaw slopes in the upper energy 
range and predicts particular correlations between the three parameters
$E$, $P$, and $T$, 
$$
	\begin{array}{ll}
	E &\propto P^2 \\
	E &\propto T^2 \\
	T &\propto P   \\
	\end{array}
	\eqno(13)
$$
while the powerlaw slopes are related to each other by
$$
	\begin{array}{ll}
	\alpha_P = 1 + {\tau_G / \tau_S} \\
	\alpha_T = \alpha_P           \\
	\alpha_E = (\alpha_P+1)/2     \\
	\end{array}
	\eqno(14)
$$
How does this basic model compare with the observations of solar flares?
The frequency distribution that is closest to a perfect powerlaw function
is that of the peak counts $P$, which has a slope of $\alpha_P=1.72\pm0.08$
(Paper I; Figure 1)
and thus constrains the ratio of the growth time to the mean saturation
time, \ie, $(\tau_G/\tau_S) = \alpha_P - 1 = 0.72$. For the slope
of flare durations we observed $\alpha_T=1.98\pm0.35$ (Paper I; Figure 3), 
which somewhat
disagrees with the theoretical expectation $\alpha_T=\alpha_P$ by about 
15\%. For the slope of total counts we observed $\alpha_E=1.60\pm0.14$
(Paper I; Figure 2),
while our model predicts $\alpha_E=(\alpha_P+1)/2 \approx 1.36$, deviating
by about 15\%. The model implies also a sharp correlation between the
parameters $P, E$, and $T$ (Equation 14), which of course is an idealization 
that is not realistic, as the scatterplots between the parameters demonstrate
in Figure 4 (of Paper I). A more realistic model can be constructed by 
introducing a
distribution of values for the instability threshold level $W_0$ and the 
decay time constant $\tau_D$, which are assumed to be constant 
(or $\delta$-functions) in our basic model. However, our basic model 
can be treated analytically and yields physical insights into SOC 
distributions. A generalization to a more realistic model (in the next
Section) requires numerical simulations. 

\begin{figure}
\centerline{\includegraphics[width=1.0\textwidth]{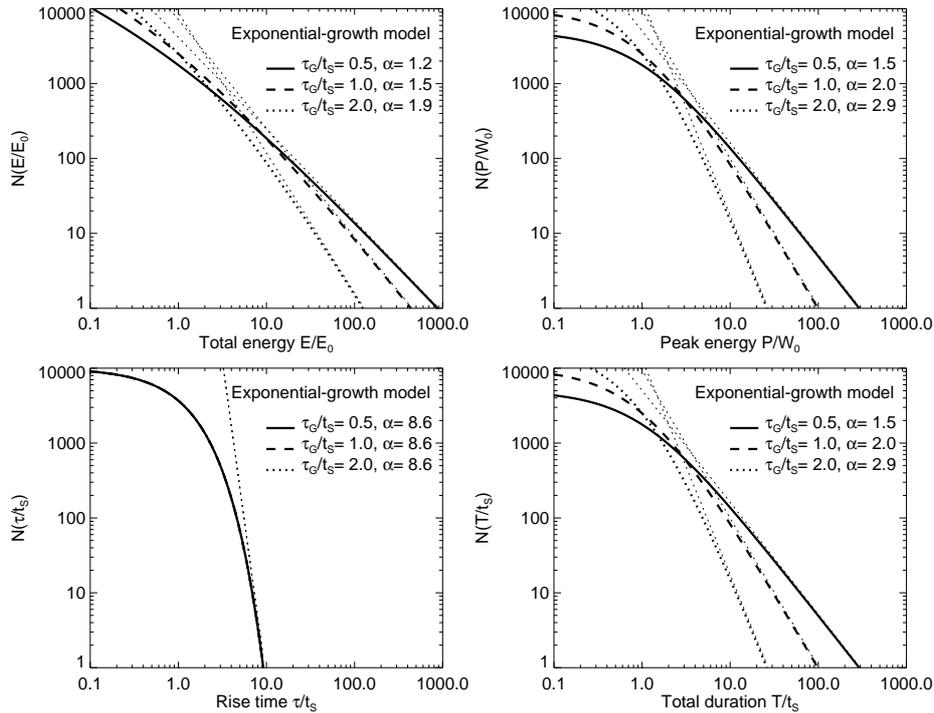}}
\caption{Frequency distribution of total energies $E$ (top left),
peak energies $P$ (top right), energy release times $\tau$ (bottom left),
and total time durations $T$ (bottom right) for the exponential-growth 
model, for $\tau_G/\tau_S$=0.5, 1, and 2. Powerlaw fits are performed at the 
upper end of the distributions (dotted thin lines), with the slopes $\alpha_P$ 
indicated in each panel. The distributions are normalized to $N_0=10^4$
events.}
\end{figure}

\section{Randomized Exponential-Growth SOC Model}

In addition to the three conditions of our basic {\sl exponential-growth
model} (Section 2), we add now also a randomization of instability
threshold energies $W_1$ (with a mean of $W_0$), which might better 
reproduce the scatter between observed parameters. It is a necessary 
condition that the critical threshold of a SOC system contains a large
number of metastable states that are close to becoming unstable,
which may be best described by some sort of a random distribution,
which we have to choose empirically due to the lack of a comprehensive
SOC theory. For a random distribution of threshold energies $W_1$ 
we may assume a normal or a log-normal (Gaussian) distribution. 
Since a normal distribution with a positive mean energy $W_0$ contains 
also negative values $W_1 < 0$, which are unphysical in our model, 
we choose a log-normal distribution $N(W_1)$, which contains by 
definition only positive values, 
$$
	N(W_1) \propto \exp{\left( - {[log(W_1)-log(W_0)]^2 
	\over 2 \sigma_W^2 }\right)} .
	\eqno(15)
$$
Drawing threshold values $W_1$ from such a log-normal distribution
and random saturation times $t_S$ from a random Poisson distribution
(approximated by an exponential function, Equation 3),
the resulting saturation energies $W_S$ have also a large spread,
$$
	W_S = W_1 \exp{ \left( {t_S \over \tau_G} \right) } . 
	\eqno(16)
$$
In addition, we add also a randomization of decay times $t_D$,
parameterized with a Poisson distribution and approximated by a
normalized exponential distribution,
$$
        N(t_D) d t_D = {N_0 \over \tau_D}
	\exp{\left(-{t_D \over \tau_D}\right)} d t_D \ .
        \eqno(17)
$$
where $t_D$ are individual decay times and $\tau_D$ is the e-folding
time constant. The physical units of the variables $\tau_G,
t_S,$ and $t_D$ are all time units (\ie, seconds).

\begin{figure}
\centerline{\includegraphics[width=1.0\textwidth]{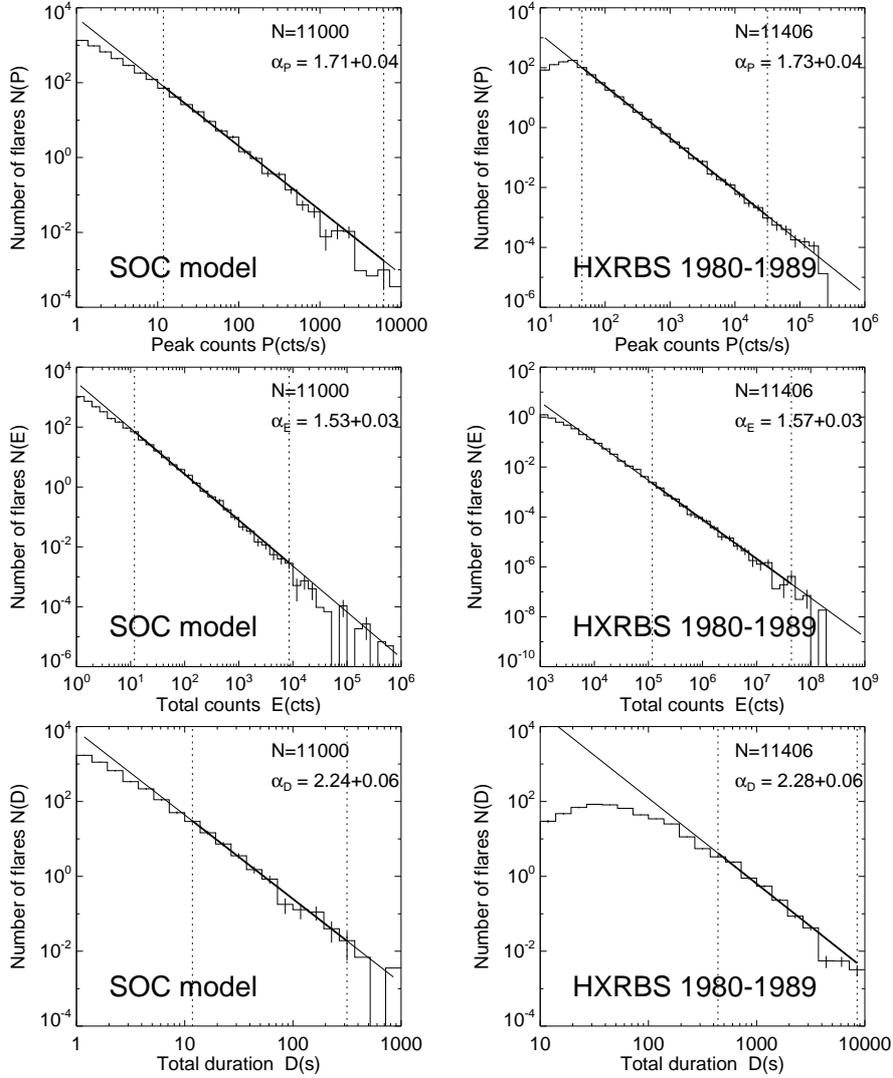}}
\caption{Frequency distribution of peak counts $P$ (top panels), 
total counts $E$ (middle panels), and event durations $D$ (bottom panels),
simulated with the randomized exponential-growth SOC model 
($\tau_S=1, \tau_G=1.5, \tau_D=1.0, W_0=4, \sigma_W=1.5$) for a 
dataset of $N=11,000$ events (left side), and compared with
HXRBS/SMM observations (right side). The weighted linear regression
fits are indicated with a thick line in the fitted range (bracketed
by dotted lines).}
\end{figure}

\begin{figure}
\centerline{\includegraphics[width=1.0\textwidth]{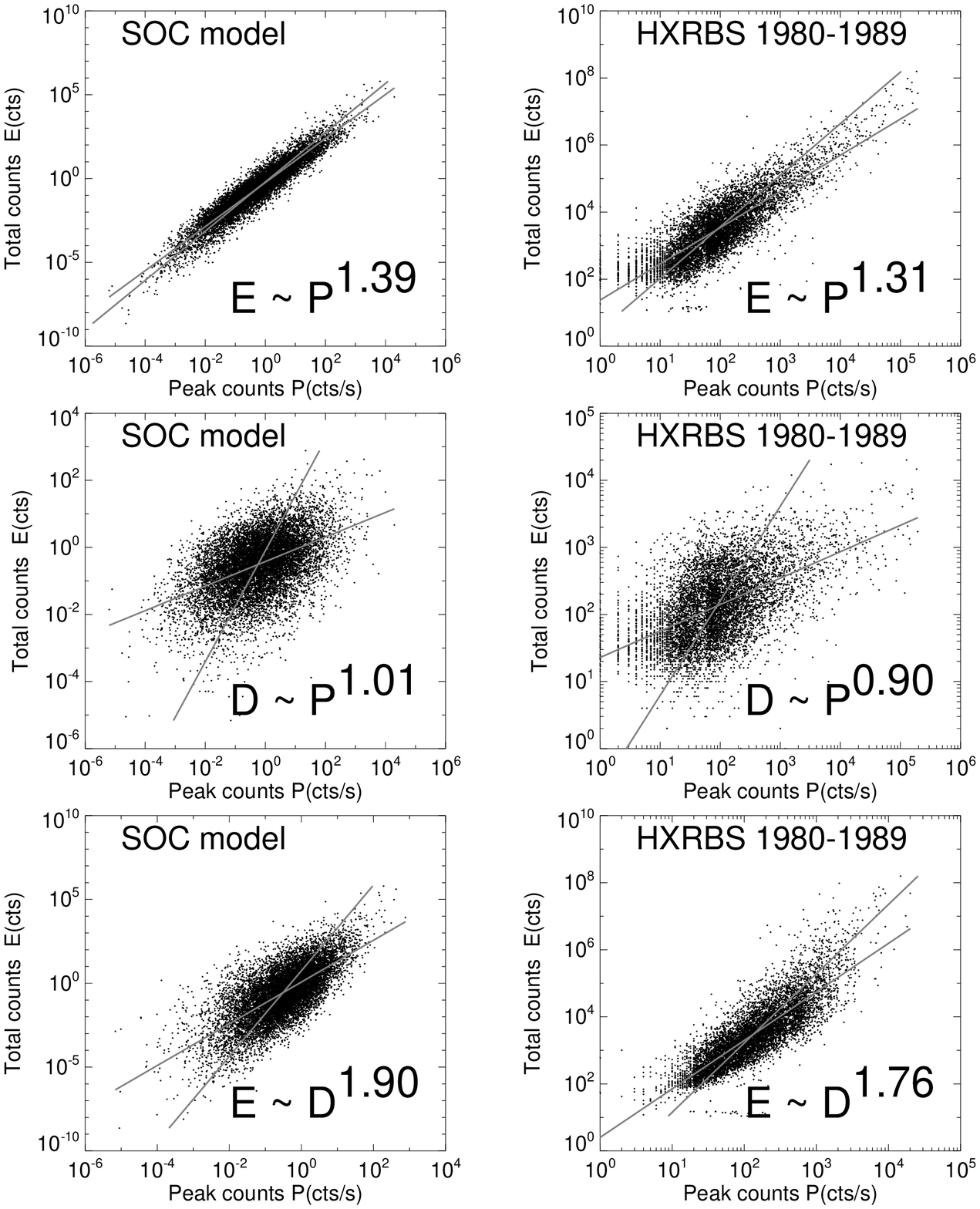}}
\caption{Scatterplots between the peak count rates $P$,
total counts $E$, and flare durations $D$ (left panels)
for the same Monte-Carlo simulations of Figure 3, compared
with observed values from HXRBS/SMM (1980 -- 1989) (right panels).
Note the similar scaling laws for the correlations between the
parameters.}
\end{figure}

Based on the random distributions of the values for
threshold energies $W_1$ (obtained from Equation 15), saturation times
$t_S$ (Equation 3), and decay times $t_D$ (Equation 17), we can then 
obtain for each parameter combination the peak counts $P$ (with Equation 2),
$$
	P = W_0 \left[ \exp{ \left( { t_S \over \tau_G} \right)} - 1 \right]  \ ,
	\eqno(18)
$$
the decay time $D$ (with Equation 6),
$$
	D = t_D \left( {P \over W_0} \right) \ ,
	\eqno(19)
$$
the total flare duration (Equation 8),
$$
	T = t_S + D \ ,
	\eqno(20)
$$
and the total counts $E$ (with Equation 10 and Figure 1),
$$
	E = P \tau_G - W_0 t_S + {1 \over 2} P D \ ,
	\eqno(21)
$$
with the reference value $E_0 = W_0 \tau_G$. 

We perform a Monte-Carlo simulation for a set of $N=11,000$ 
events, which mimic the distributions of solar flares averaged over a 
whole solar cycle as observed with HXRBS/SMM during 1980--1989 
(Figures 1--3 in Paper I). 
For each event with randomized sets of values $(t_S, \tau_G, t_D)$
according to the distributions defined in Equations (2), (15), (17) we
calculate the parameters $P$, $T$, and $E$ (according to Equations 18, 19,
and 20) and show the resulting distributions in Figure 4 (left-hand panels).
We obtain a good match with the observed distributions (Figure 4, righ-hand
panels, or Figures 1 -- 3 in Paper I), with slopes of 
$\alpha_P \approx 1.71\pm0.04$ (vs. RHESSI: $1.73\pm0.04$), 
$\alpha_E \approx 1.53\pm0.03$ (vs. RHESSI: $1.57\pm0.03$), and
$\alpha_T \approx 2.24\pm0.06$ (vs. RHESSI: $2.28\pm0.06$),
for the following model constants: 
$\tau_S=1.0$, $\tau_G=1.5$, $\tau_D=1.0$, $W_0=4$, and $\sigma_W=1.5$. 
Note also that the simulated frequency distributions $N(P), N(E)$, and $N(T)$ 
all exhibit a roll-over at the lower end of the distributions. 

The scatterplots between the three
variables $P$, $E$, and $T$ are shown in Figure 4,
which also display a comparable scatter as observed in the data, 
with similar linear regression fits as observed, i.e., 
$E \propto P^{1.39}$ (vs. RHESSI: $E \propto P^{1.31}$),
$D \propto P^{1.01}$ (vs. RHESSI: $D \propto P^{0.90}$), and 
$E \propto D^{1.90}$ (vs. RHESSI: $E \propto D^{1.76}$). 

Thus, we conclude that this model provides 
more realistic occurrence frequency distributions and parameter
correlations than the idealized model with a fixed threshold level 
$W_0$ and decay time $\tau_D$.  

\begin{figure}
\centerline{\includegraphics[width=1.0\textwidth]{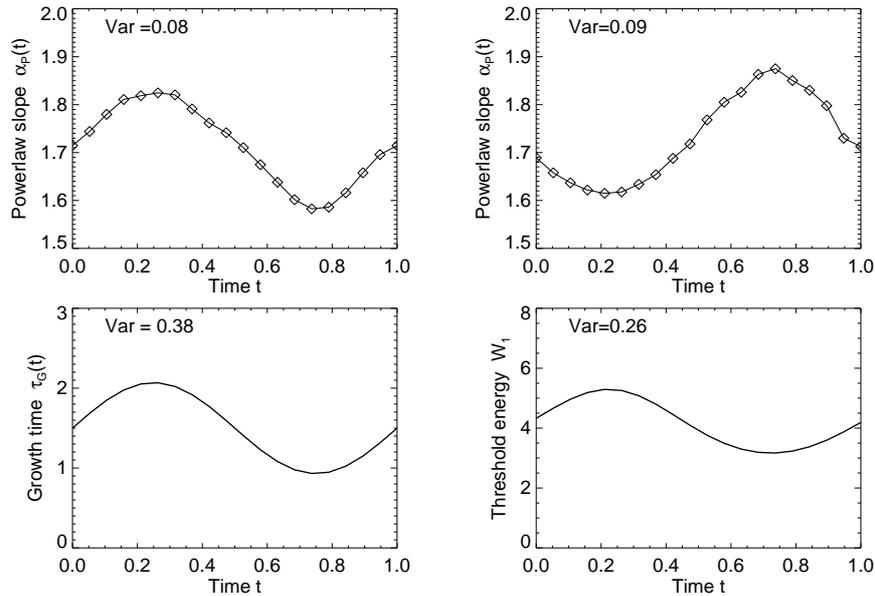}}
\caption{Monte-Carlo simulations of a variation of the powerlaw
slope of flare peak fluxes by 8\% (top), which can be caused by
a variation of the growth time $\tau_G$ by 38\% (bottom left)
or by a variation of the mean threshold energy by 26\%
(bottom right).}
\end{figure}

\section{Discussion and Conclusions}

The paramount interpretation of the powerlaw shape of frequency
distributions of solar flare parameters is the state of self-organized 
criticality of a dissipative system (Lu and Hamilton, 1991). This
model predicts scale-free distributions of the size parameters of
flare events, such as temporal, spatial, and energy parameters.
We measured durations, peak values, and time-integrated values of
hard X-ray photon counts, which serve as a good proxy of the flare
duration, peak energy release rate, and total released flare energies,
based on the thick-target bremsstrahlung flare model (Brown, 1971) and
the correlations found between these parameters (Crosby \etal 1993). 
The detailed functional shape of the resulting frequency
distributions and the scatterplots of the correlated parameters have
been modeled with cellular automaton models before 
(\eg Lu and Hamilton etal 1991). 

A basic model based on multiplicative nonlinear
processes (with exponential growth), random saturation times, and
linear decay (Section 2) predicts an exact powerlaw function at 
the upper end of all frequency distributions, with a rollover at 
the lower end (Figure 2). However, the predicted relationships between 
the powerlaw slopes, $\alpha_P=\alpha_T$ and $\alpha_E=(\alpha_P+1)/2$,
do not agree with the observations (\ie, $\lapprox 20\%$) within the
accuracy of the measured powerlaw slopes ($\lapprox 5\%$). Also the
predicted correlation between the parameters does not allow for any
scatter, in strong contrast to the observed datapoints. 

We develop a more realistic SOC model with randomized parameters (of the
threshold levels, saturation times, and decay times), which is able to
reproduce closely the powerlaw slopes of the observed frequency distributions 
and the scatter between correlated parameters.
This randomized exponential-growth SOC model predicts different
relationships between the powerlaw slopes than the idealized one,
which cannot be calculated analytically without approximations, but
can easily be generated with numerical (Monte-Carlo) simulations.

In Paper I we discovered a systematic modulation of the powerlaw slope
of the frequency distributions of solar flares during three solar cycles, 
which implies also a modulation of the physical conditions in solar
flare sites. A possible coupling could occur between the solar dynamo 
and the magnetic complexity in the solar photosphere, chromosphere, 
and corona that leads to magnetic instabilities resulting into flares.
The detailed relationship between magnetic complexity and multiplicative 
chain reactions that occur in a SOC system depends on specific physical 
models such as magnetic reconnection scenarios. In terms of our analytical
SOC model, the powerlaw slope of flare frequency distributions can be
modulated by the ratio of growth to saturation times ($\alpha_P
=1+\tau_G/\tau_S$; Eq.~5), which can be driven by longer growth times
$\tau_G$ in magnetically more complex regions during high solar activity
periods. A Monte-Carlo simulation shows that a variation of the mean
instability growth time by 38\% is required to induce a modulation of
the powerlaw slope by 8\% (Figure 5 left), as observed during the last 
three solar cycles (Paper I). A shorter growth rate corresponds to 
faster nonlinear evolution, which could occur in reconnection regions
with higher magnetic stress. 
Alternatively, the mean threshold energy $W_1$ could be larger 
in magnetically unstable regions during high solar activity periods.
A Monte-Carlo simulation shows that a change of the mean threshold
energy by 26\% can explain the observed powerlaw slope variation
of 8\% (Figure 5 top). A higher threshold could be obtained by stronger
magnetic stressing during the solar cycle maximum.
Thus, detailed forward-fitting of our analytical SOC model to the
observed distributions in various active regions and in specific
time intervals of the solar cycle could diagnose the physical conditions
of magnetic configurations, which provides important information for 
statistical flare forecasting.

\acknowledgements 
We thank the referee Brian Dennis for constructive and helpful 
comments. This work is partially supported by NASA contract
NAS5--98033 of the RHESSI mission through the University of California,
Berkeley (subcontract SA2241--26308PG) and NASA grant NNX08AJ18G.
We acknowledge access to solar mission data and flare catalogs from the
{\sl Solar Data Analysis Center} (SDAC) at the NASA Goddard Space Flight
Center (GSFC).

\section*{References} 

\def\ref#1{\par\noindent\hangindent1cm {#1}}
\small
\ref Aschwanden, M.J., Dennis, B.R., Benz, A.O. 1998,
        Logistic avalanche processes, elementary time structures,
        and frequency distributions of flares,
        {\it Astrophys. J.} {\bf 497}, 972-993.
\ref Aschwanden, M.J., 2010a,
        The State of Self-Organized Criticality of the Sun
        During the Last Three Solar Cycles. I. Observations, 
        Solar Physics, (this volume), subm. (Paper I).
\ref Aschwanden, M.J., 2010b,
	{\sl Self-Organized Criticality in Astrophysics.
	The Statistics of Nonlinear Processes in the Universe},
	Springer-Praxis: Heidelberg, New York (in press). 
\ref Bak, P., Tang, C., Wiesenfeld, K. 1987,
 	Self-organized criticality: An explanation of 1/f noise,
 	{\it Phys. Rev. Lett.} {\bf 59/4}, 381-384.
\ref Brown, J.C. 1971,
        The deduction of energy spectra of non-thermal electrons
        in flares from the observed dynamic spectra of Hard X-Ray bursts,
        {\it Solar Phys.} {\bf 18}, 489-502.
\ref Charbonneau, P., McIntosh, S.W., Liu, H.L., and Bogdan, T.J. 2001,
 	Avalanche models for solar flares,
 	{\it Solar Phys.}, {\bf 203}, 321-353. 
\ref Crosby, N.B., Aschwanden, M.J., Dennis, B.R. 1993,
        Frequency distributions and correlations of solar X-ray
        flare parameters, {\it Solar Phys.} {\bf 143}, 275-299.
\ref Crosby, N.B. 1996,
        {\sl Contribution \`a l'Etude des Ph\'enom\`enes Eruptifs
        du Soleil en Rayons Z \`a partir des Observations de l'Exp\'erience
        WATCH sur le Satellite GRANAT},
        PhD Thesis, University Paris VII, Meudon, Paris. 
\ref Fermi, E. 1949, On the origin of the cosmic radiation,
	{\it Phys. Rev. Lett.} {\bf 75}, 1169.
\ref Georgoulis, M.K., Vilmer,N., Crosby,N.B. 2001,
        A Comparison Between Statistical Properties of Solar X-Ray
        Flares and Avalanche Predictions in Cellular Automata Statistical
        Flare Models, {\it Astron. Astrophys.} {\bf 367}, 326-338.
\ref Huberman, B.A. and Adamic, L. 1999,
	Growth dynamics of the World-Wide Web,
	{\it Nature} {\bf 401}, 131. 
\ref Lu, E.T. and Hamilton, R.J. 1991,
 	Avalanches and the distribution of solar flares,
 	{\it Astrophys. J.}, {\bf 380}, L89-L92.
\ref Lu, E.T. 1995,
        Constraints on energy storage and release models for astrophysical
        transients and solar flares, 
	{\it Astrophys. J.}, {\bf 447}, 416-418.
\ref Rosner, R. and Vaiana, G.S. 1978,
	Cosmic flare transients: constraints upon models for energy
	storage and release derived from the event frequency distribution,
	{\it Astrophys. J.} {\bf 222}, 1104-1108.
\ref Simkin, M.V., and Roychowdhury, V.P. 2006,
 	Re-inventing Willis, eprint arXiv:physics 0601192. 
\ref Wheatland, M.S. 2000,
        Do solar flares exhibit an interval-size relationship?
        {\it Solar Phys.}, {\bf 191}, 381-389.
\ref Willis, J.C. and Yule, G.U. 1922,
	Some statistics of evolution and geographical distribution
	in plants and animals, and their significance,
	{\it Nature}, {\bf 109}, 177-179.

\end{article}
\end{document}